\documentclass[aip,jcp,reprint,longbibliography]{revtex4-2}

\usepackage{graphicx}
\usepackage{bm}
\usepackage{physics}
\usepackage{mathtools}
\usepackage{amssymb}
\usepackage{amsthm}
\usepackage{mathrsfs}
\usepackage{tensor}
\usepackage{xcolor}
\usepackage[hidelinks]{hyperref}
\usepackage{booktabs}
\usepackage{multirow}
\usepackage{enumitem}

\begin{document}

\title{State--Generator Geometry of Open Quantum Systems: Compatibility and Covariant Transport}

\author{Eric R. Bittner}
\email{ebittner@central.uh.edu}
\affiliation{Department of Physics, University of Houston, Houston, Texas 77204, USA}
\affiliation{Institut Courtois \& D\'epartement de physique, Universit\'e de Montr\'eal, 1375 Avenue Th\'er\`ese-Lavoie-Roux, Montr\'eal H2V~0B3, Qu\'ebec, Canada}

\author{Carlos~Silva-Acu\~na}
\email{carlos.silva@umontreal.ca}
\affiliation{School of Chemistry and Biochemistry, Georgia Institute of Technology, 901 Atlantic Drive, Atlanta, GA~30332, United~States}
\affiliation{Institut Courtois \& D\'epartement de physique, Universit\'e de Montr\'eal, 1375 Avenue Th\'er\`ese-Lavoie-Roux, Montr\'eal, Qu\'ebec H2V~0B3, Canada}
\affiliation{Departamento de F\'isica Aplicada, Centro de Investigaci\'on y de Estudios Avanzados del Instituto Polit\'ecnico Nacional (CINVESTAV), 97310 M\'erida, Yucat\'an, M\'exico}

\date{\today}

\begin{abstract}
We develop a geometry for transporting stationary-state response across the
control space of an open quantum system.  A physical model is represented by
the ordered pair of its stationary state and dynamical generator.  Embedding
these pairs in a common ambient space induces a metric, a response one-form,
and a closed two-form on the control manifold.  The ambient space admits a
canonical complex structure that exchanges state and generator directions,
but the Liouvillian null-state condition restricts physical models to a
submanifold that need not preserve this structure.  For an amplitude-damped
optical Bloch model, the metric and response two-form are compatible at the
single point $\omega=0$ and $g/\Gamma=1/\sqrt{2}$; the physical manifold is
non-K\"ahler elsewhere.  The induced metric defines the Levi--Civita
connection, geodesics, and parallel transport without requiring K\"ahler
compatibility.  We compute the connection by automatic differentiation
through the stationary Liouvillian solve and recover the symbolic result to
machine precision.  A two-point calculation then shows that the resulting
geodesic differs from linear interpolation in control space and follows a
shorter path through the family of state--generator models.  This geometry
supplies the intrinsic derivative and transport structure needed to carry
observable response, including multidimensional spectra, between admissible
stationary models.

\end{abstract}

\maketitle

\section{Introduction}
\label{sec:introduction}

Geometry provides a common language for describing how physical systems
respond to changes in their control parameters. In equilibrium
thermodynamics, the Hessian geometries of Weinhold and Ruppeiner organize
fluctuations, susceptibilities, stability, and critical behavior
\cite{Weinhold1975,Ruppeiner1995}. In quantum mechanics, Berry curvature,
the quantum geometric tensor, and information metrics characterize phase,
transport, distinguishability, and parameter sensitivity
\cite{Berry1984,Provost1980,Zanardi2007,Gu2010}. These developments share a
basic idea: a family of physical models can be regarded as a manifold, and
response to parameter variation can be encoded in geometric objects defined
on that manifold.

The geometry of response becomes richer away from equilibrium. Equilibrium
response derives from a thermodynamic potential and obeys reciprocity, so its
local structure is symmetric and metric. A nonequilibrium stationary state
need not derive from a scalar potential, and its response tensor can possess
an independent antisymmetric sector. In our earlier geometric formulation of
open-system thermodynamics, quasistatic evolution follows a manifold of
stationary states and the work performed around a control-space cycle is the
flux of a response curvature \cite{Bittner:2026aa}. In our subsequent work, we
showed that the symmetric sector defines a susceptibility metric while the
antisymmetric sector is the curvature two-form governing nonreciprocal
response and geometric work \cite{JCP2}. Open quantum systems can carry metric
and symplectic sectors on the same control manifold.

That result raises a structural question. A metric and a closed two-form are
the central ingredients of an almost K\"ahler geometry, but their coexistence
does not by itself establish compatibility. One must also identify an
almost-complex structure \(J\), satisfying \(J^2=-I\), that relates the two
sectors through
\[
F(X,Y)=g(JX,Y).
\]
For a genuine K\"ahler structure, \(J\) must additionally preserve the
physical tangent spaces and be integrable. The response decomposition of the
preceding paper suggested that such a completion might exist, while leaving
both the origin of \(J\) and its physical meaning unresolved.

The missing structure becomes natural when a quantum dynamical model is
represented not by its stationary state alone, nor by its generator alone,
but by the ordered pair of the two. We call this the \emph{state--generator
representation}. A parameterized family of models then defines an embedding
\[
\Phi:\mathcal M\longrightarrow\mathcal E,
\qquad
\Phi(\boldsymbol\lambda)
=
\bigl(\mathbf r(\boldsymbol\lambda),
      \mathbf h(\boldsymbol\lambda)\bigr),
\]
from the control manifold \(\mathcal M\) into an ambient state--generator
space \(\mathcal E=\mathcal R\oplus\mathcal H\). Tangent vectors to this
embedding contain simultaneous variations of the state and its dynamical
generator. The embedding carries more information than the
stationary-state manifold by itself: it records both how the system changes
and how the law generating its dynamics changes.

The same representation offers a basis for comparing open-system models.
Distinct generators can share the same stationary state, so a geometry built
from stationary states alone cannot distinguish their dynamics.  The joint
embedding separates such models and assigns a local measure of dissimilarity
through the induced metric.  Geodesic distance, curvature, the rank of the
embedding, and the compatibility of the metric and response two-form then
provide coordinate-independent signatures with which families of models may
be organized or classified, once a common operator basis and normalization
have been fixed.

We restrict the present theory to finite-dimensional, time-homogeneous
Markovian dynamics generated by a Liouvillian of
Gorini--Kossakowski--Sudarshan--Lindblad form
\cite{Gorini1976,Lindblad1976,Breuer2002}.  In this setting, a
stationary state is a null state of
$\mathcal L(\boldsymbol\lambda)$, and the pair
$(\rho_{\rm ss},\mathcal L)$ is specified by one point in control space.  The
paths considered below run through this space of stationary models; their path
parameter is not physical time.  If such a path is realized by slowly varying
experimental controls, the intended limit is quasistationary: the system
relaxes to the stationary state at each point and retains no temporal memory
of the parameter-space transport.  The Markovian assumption fixes the model
descriptor used here, rather than the meaning of transport itself.  We return
to possible non-Markovian model descriptors in Sec.~\ref{sec:discussion}.

This paired representation supplies a canonical geometry. The
ambient inner product induces a metric on \(\mathcal M\), while the natural
state--generator pairing defines a response one-form and its closed two-form.
Because the state and generator sectors enter as complementary components,
the ambient space also admits a canonical quarter-turn that exchanges them.
This map is the sought-after complex structure: it converts state variations
into generator variations and relates the symmetric and antisymmetric response
sectors. The remaining issue is whether it restricts consistently to the
embedded manifold of physically admissible state--generator pairs.

The distinction between ambient compatibility and physical compatibility is
essential. The canonical complex structure is automatic in the full direct
sum \(\mathcal E\), but the equations of motion constrain the physical image
\(\Phi(\mathcal M)\) to a submanifold. An almost K\"ahler structure is inherited
only when the canonical map preserves its tangent spaces. A K\"ahler structure
requires, in addition, vanishing Nijenhuis tensor. These conditions turn the
open question posed by the response decomposition into concrete tests on the
state--generator embedding.

Compatibility also determines how response objects should be compared at
neighboring models. The induced metric defines the unique torsion-free,
metric-compatible Levi--Civita connection, which removes changes of the local
tangent basis from parameter derivatives. Covariant differentiation and
parallel transport then provide an intrinsic means of carrying response data
through model space. The state--generator geometry connects
the local response geometry of the preceding paper to a subsequent theory of
Liouvillian response transport. Explicit transport expansions and their
application to nonlinear spectroscopic pathways belong to the companion work
\cite{JCP3}.

The present paper has three aims. First, we formulate
the state--generator representation and derive the metric, response potential,
and closed two-form induced on the control manifold. Second, we construct the
canonical complex structure and state the tangency and integrability
conditions under which the physical manifold becomes almost K\"ahler or
K\"ahler. Third, we derive the associated Levi--Civita connection, covariant
derivative hierarchy, and parallel-transport law. The result is a geometric
bridge between stationary-state response and Liouvillian transport that is
independent of any particular numerical implementation.

The remainder of the paper is organized as follows. Section~\ref{sec:II}
introduces the state--generator embedding and its induced response geometry.
Section~\ref{sec:complex-structure} develops the canonical complex structure
and the compatibility of \(g\) and \(F\). Section~\ref{sec:covariant-transport}
constructs the Levi--Civita connection, covariant differentiation, and
parallel transport. Section~\ref{sec:examples} discusses illustrative models,
and Sec.~\ref{sec:discussion} examines the scope and implications of the
theory.

\section{Foundational Principles of Response Geometry}
\label{sec:II}

The response metric and curvature introduced in the preceding paper coexist
on the control manifold, but their compatibility requires a representation
that treats the physical state and its dynamical generator together. We seek
a description in which neighboring physical models
are points of an embedded state--generator manifold. Tangent vectors of this
embedding record simultaneous state and generator variations and provide the
objects from which metric, symplectic, and complex structures can be induced.

This section constructs the state--generator representation and its induced
response geometry. The resulting structure supplies the intrinsic language
needed to compare neighboring models and prepares the covariant transport
theory developed later in the manuscript.

\subsection{Principle I: State--Generator Representation}

A quantum dynamical model is completely characterized by its physical state and the generator governing its evolution.
We refer to this pair
as the state--generator representation. 
Further, every quantum dynamical model depends upon a set of experimentally or
theoretically controllable parameters,
\[
\boldsymbol{\lambda}
=
(\lambda^1,\lambda^2,\ldots,\lambda^N),
\]
including Hamiltonian parameters (site energies, electronic
couplings, external fields), dissipative parameters (relaxation and
dephasing rates), environmental variables (temperature, spectral
densities), and quantities associated with state preparation and
measurement. Collectively, these parameters specify the physical
model.
The geometric structures developed in this work are induced
from this representation, while the transport laws and computational
tools arise from its differential structure.

The collection of all admissible parameter sets defines an
$N$-dimensional manifold of physical models,
$\mathcal M$,
whose points are labeled by
\(
\boldsymbol{\lambda}\in\mathcal M.
\)
Each point corresponds to a unique quantum dynamical model, while
neighboring points represent infinitesimal variations of that model,
\[
d\boldsymbol{\lambda}
=
(d\lambda^1,\ldots,d\lambda^N).
\]
Rather than treating neighboring models as independent calculations,
we regard them as neighboring points on a differentiable manifold and
seek a geometry that relates their observables through transport.

To each point $\boldsymbol{\lambda}\in\mathcal M$
we associate an ordered pair consisting of the physical state and the
generator governing its evolution,
$
(\mathbf r,\mathbf h).
$
The state vector $\mathbf r$ and generator vector $\mathbf h$ reside in
the ambient state--generator space
$
\mathcal E
=
\mathcal R
\oplus
\mathcal H,
$
where $\mathcal R$ and $\mathcal H$ denote the state and generator
subspaces, respectively. The state--generator map introduced below
embeds the manifold of physical models into this ambient space.

For a finite-dimensional quantum system represented in an operator
basis containing
$
N_{\mathrm{op}}
$
linearly independent generators,
\[
\mathcal R
\simeq
\mathbb R^{N_{\mathrm{op}}},
\qquad
\mathcal H
\simeq
\mathbb R^{N_{\mathrm{op}}},
\]
so that
$
\mathcal E
\simeq
\mathbb R^{2N_{\mathrm{op}}}.
$

In particular, when the operator basis is chosen from the Lie algebra
$
\mathfrak{su}(N),
$
the number of generators is
$
N_{\mathrm{op}}
=
N^2-1,
$
giving
$
\mathcal E
\simeq
\mathbb R^{2(N^2-1)}.
$
The dimension of the ambient space depends only on the chosen operator
representation and plays no role in the geometry.
The smooth embedding representing the physical model is
$
\Phi:
\mathcal M
\longrightarrow
\mathcal E,
$
defined by
\[
\Phi(\boldsymbol{\lambda})
=
\left(
\mathbf r(\boldsymbol{\lambda}),
\mathbf h(\boldsymbol{\lambda})
\right).
\]
The vector
$
\mathbf r
$
contains the coordinates of the physical state in the chosen operator
basis, while
$\mathbf h$
contains the coordinates of the corresponding dynamical generator.
The ordered pair
$(\mathbf r,\mathbf h)$
will be referred to as the
\emph{state--generator representation}.  It constitutes the
fundamental object of the present theory.  All subsequent geometric
structures are induced from this mapping.
The state--generator representation is independent of the particular
operator basis used to represent the dynamics.

For open quantum systems, the state--generator representation is
constructed from the stationary solution of the Liouvillian dynamics.
At each point
\(
\boldsymbol{\lambda}\in\mathcal M,
\)
the steady-state density operator satisfies
\[
\mathcal L(\boldsymbol{\lambda})
\rho_{\mathrm{ss}}(\boldsymbol{\lambda})
=
0,
\]
where
\(
\mathcal L
\)
denotes the Liouvillian superoperator. The state vector
\(
\mathbf r(\boldsymbol{\lambda})
\)
is the representation of
\(
\rho_{\mathrm{ss}}
\)
in the chosen operator basis, while
\(
\mathbf h(\boldsymbol{\lambda})
\)
contains the coordinates specifying the corresponding dynamical
generator. Each point of the state--generator manifold
represents a stationary state together with the generator that
produces it.

The embedding
\(
\Phi:\mathcal M\rightarrow\mathcal E
\)
is constrained by the equations of motion. Its image is not
the full Cartesian product
\[
\mathcal R\times\mathcal H,
\]
but the submanifold of physically admissible state--generator pairs
satisfying
\[
\mathcal L(\mathbf h)\,\rho_{\mathrm{ss}}=0.
\]
The geometry developed below is the intrinsic geometry of this
embedded solution manifold rather than that of the ambient
state--generator space.

\subsection{Principle II: Induced Response Geometry}

Representing a quantum dynamical model by the state--generator map
\[
\Phi:\mathcal M\rightarrow\mathcal E
\]
raises a natural question: what geometric structure does this
representation carry?
The geometry is inherited from the embedding rather than imposed
independently.  Infinitesimal changes of the state and generator define
the tangent basis
\[
\mathbf e_\mu
=
\partial_\mu\Phi
=
\left(
\partial_\mu\mathbf r,
\partial_\mu\mathbf h
\right),
\]
where
\[
\partial_\mu
=
\frac{\partial}{\partial\lambda^\mu}.
\]
Each tangent vector represents the response of the state and its
generator to an infinitesimal variation of the model parameters.
We equip the ambient space
\[
\mathcal E
=
\mathcal R
\oplus
\mathcal H
\]
with the canonical Euclidean inner product
\[
\left<
(\mathbf a,\mathbf b),
(\mathbf c,\mathbf d)
\right>
=
\mathbf a\cdot\mathbf c
+
\mathbf b\cdot\mathbf d.
\]
The metric is then induced by pullback,
\[
g
=
\Phi^\ast
g^{(\mathcal E)},
\]
which, in local coordinates, becomes
\[
g_{\mu\nu}
=
g^{(\mathcal E)}
(\mathbf e_\mu,\mathbf e_\nu)
=
\partial_\mu\mathbf r
\cdot
\partial_\nu\mathbf r
+
\partial_\mu\mathbf h
\cdot
\partial_\nu\mathbf h.
\]
Since $g_{\mu\nu}$ is the Gram matrix of the tangent vectors,
\[
g_{\mu\nu}
=
\mathbf e_\mu\cdot\mathbf e_\nu,
\]
the following properties are immediate:
\begin{enumerate}
\item
$g_{\mu\nu}=g_{\nu\mu}$,
\item
$g$ is positive semi-definite,
\item
$g$ is positive definite whenever the tangent vectors are linearly
independent.
\end{enumerate}
The parameter manifold inherits a Riemannian structure
whenever the state--generator embedding is regular.

The metric measures the distinguishability of neighboring physical
models through simultaneous variations of both the quantum state and
its dynamical generator.  Unlike the response metric introduced in
previous work, the present metric arises directly from the geometry of
the embedding and is independent of any particular observable.

The state--generator representation also possesses a canonical
bilinear pairing,
\[
A
=
\mathbf r\cdot d\mathbf h,
\]
which defines a differential one-form on the parameter manifold.

Expressed in local coordinates,
\[
A
=
A_\mu
d\lambda^\mu,
\]
where
\[
A_\mu
=
\mathbf r
\cdot
\partial_\mu\mathbf h.
\]
The one-form $A$ defines the  response potential, coupling the
instantaneous physical state to infinitesimal variations of the
dynamical generator.

Applying the exterior derivative gives the closed two-form
\[
F = dA
= d\mathbf r
\wedge
d\mathbf h,
\]
whose coordinate representation is
\[
F_{\mu\nu}
=
\partial_\mu\mathbf r
\cdot
\partial_\nu\mathbf h
-
\partial_\nu\mathbf r
\cdot
\partial_\mu\mathbf h.
\]

The antisymmetry,
\[
F_{\mu\nu}
=
-
F_{\nu\mu},
\]
follows immediately from the properties of the exterior product, while
\[
dF
=
0
\]
shows that the response two-form is closed.

The tensor $F$ measures the oriented state--generator area swept out
by simultaneous parameter variations and characterizes the
circulatory component of response.  In the present formulation it
arises naturally from the canonical pairing between state and
generator and coincides with the antisymmetric response tensor
introduced in our previous work.

The same state--generator map induces the metric $g$, response potential $A$,
and closed two-form $F$.  Whether the
symmetric and antisymmetric sectors define a compatible complex
geometry is a separate question, addressed in
Sec.~\ref{sec:complex-structure}.  Once compatibility and the induced
connection are established, the geometry can be used to compare and
transport objects between neighboring physical models.

\subsection{Geometric Outlook: Toward Liouvillian Transport}
\label{sec:transport-outlook}

The state--generator embedding and its induced response geometry determine how tangent objects at neighboring physical models may be compared. The remaining step is to construct transport laws from the associated connection and to relate those laws to observable response. We develop the intrinsic connection and parallel-transport structure below; explicit Liouvillian transport expansions and their computational realization are reserved for the companion manuscript.

\section{Canonical Complex Structure and Compatibility of
\texorpdfstring{$g$}{g} and \texorpdfstring{$F$}{F}}\label{sec:complex-structure}
The state--generator embedding places the state and its dynamical
generator on equal footing in the ambient space
\[
\mathcal E
=
\mathcal R\oplus\mathcal H.
\]
At each point of the embedded manifold, a tangent vector may be written
as
\[
X
=
(X_r,X_h),
\]
where
\[
X_r
=
X^\mu\partial_\mu\mathbf r,
\qquad
X_h
=
X^\mu\partial_\mu\mathbf h.
\]
Similarly,
\[
Y
=
(Y_r,Y_h).
\]

The ambient metric introduced above is
\[
g(X,Y)
=
X_r\cdot Y_r
+
X_h\cdot Y_h,
\]
while the response two-form is
\[
F(X,Y)
=
X_r\cdot Y_h
-
X_h\cdot Y_r.
\]

We now seek a linear map
\[
J:T\mathcal E\rightarrow T\mathcal E
\]
that relates the metric and response two-form through
\[
F(X,Y)=g(JX,Y).
\]
Because $F$ pairs state variations with generator variations, the
required map must exchange the state and generator components of a
tangent vector. We define
\[
J(X_r,X_h)
=
(-X_h,X_r).
\]
In block-matrix form,
\[
J
=
\begin{pmatrix}
0&-I\\
I&0
\end{pmatrix},
\]
where $I$ is the identity on the $d$-dimensional state or generator
subspace.

Applying $J$ twice gives
\[
\begin{split}
J^2(X_r,X_h)
&=
J(-X_h,X_r)
\\
&=
(-X_r,-X_h)
\\
&=
-(X_r,X_h).
\end{split}
\]
Hence
\[
J^2=-I,
\]
and $J$ defines a complex structure on the ambient state--generator
space.

The compatibility relation follows directly. Acting with $J$ on $X$
and evaluating the metric gives
\[
\begin{split}
g(JX,Y)
&=
g\bigl((-X_h,X_r),(Y_r,Y_h)\bigr)
\\
&=
-X_h\cdot Y_r
+
X_r\cdot Y_h.
\end{split}
\]
With the two-form convention
\[
F
=
d\mathbf r\wedge d\mathbf h,
\]
we have
\[
F(X,Y)
=
X_r\cdot Y_h
-
X_h\cdot Y_r,
\]
which gives
\[
\boxed{
F(X,Y)=g(JX,Y).
}
\]

Equivalently, reversing the orientation to
\[
F
=
d\mathbf h\wedge d\mathbf r,
\]
then the same choice of $J$ gives
\[
F(X,Y)=-g(JX,Y).
\]
The sign in the compatibility relation is fixed jointly by the
orientation chosen for $F$ and the definition of $J$. In the present
work we adopt the convention
\[
F
=
d\mathbf r\wedge d\mathbf h,
\qquad
J(X_r,X_h)=(-X_h,X_r),
\]
so that
\[
F(X,Y)=g(JX,Y).
\]

The complex structure is also orthogonal with respect to the metric.
Indeed,
\[
\begin{split}
g(JX,JY)
&=
g\bigl((-X_h,X_r),(-Y_h,Y_r)\bigr)
\\
&=
X_h\cdot Y_h
+
X_r\cdot Y_r
\\
&=
g(X,Y).
\end{split}
\]
Hence
\[
g(JX,JY)=g(X,Y),
\]
and $J$ preserves lengths and angles in the ambient state--generator
space.

The antisymmetry of $F$ follows from the compatibility relation:
\[
\begin{split}
F(Y,X)
&=
g(JY,X)
\\
&=
g(X,JY)
\\
&=
-g(JX,Y)
\\
&=
-F(X,Y),
\end{split}
\]
where we used
\[
g(JX,Y)=-g(X,JY),
\]
which follows from the block form of $J$.

The metric and response two-form combine into the
Hermitian tensor
\[
\mathscr H
=
g+iF.
\]
For tangent vectors $X$ and $Y$,
\[
\mathscr H(X,Y)
=
g(X,Y)
+
i\,g(JX,Y).
\]
Its real part determines the metric structure,
\[
\operatorname{Re}\mathscr H=g,
\]
while its imaginary part gives the antisymmetric response,
\[
\operatorname{Im}\mathscr H=F.
\]

These identities hold immediately in the ambient space
$\mathcal E$. For the embedded state--generator manifold
$\Phi(\mathcal M)$ to inherit this complex structure, the action of
$J$ must preserve its tangent spaces:
\[
J\bigl(T_p\mathcal M\bigr)
\subseteq
T_p\mathcal M
\qquad
\text{for all }p\in\mathcal M.
\]
Equivalently, for every tangent vector
\[
X
=
X^\mu
\left(
\partial_\mu\mathbf r,
\partial_\mu\mathbf h
\right),
\]
there must exist coefficients
\[
(JX)^\nu
\]
such that
\[
\left(
-X^\mu\partial_\mu\mathbf h,
X^\mu\partial_\mu\mathbf r
\right)
=
(JX)^\nu
\left(
\partial_\nu\mathbf r,
\partial_\nu\mathbf h
\right).
\]
When this tangency condition is satisfied, the ambient complex
structure restricts to an almost-complex structure on the manifold of
physical models.

Because
\[
F=dA,
\]
the response two-form is closed:
\[
dF=0.
\]
The induced manifold is almost K\"ahler whenever
$J^2=-I$, $g(JX,JY)=g(X,Y)$,  $F(X,Y)=g(JX,Y)$, and $dF=0$.
A genuine K\"ahler structure additionally requires the induced
almost-complex structure to be integrable. Equivalently, its Nijenhuis
tensor must vanish,
\[
N_J(X,Y)
=
[JX,JY]
-
J[JX,Y]
-
J[X,JY]
-
[X,Y]
=
0.
\]
For the constant ambient complex structure defined above,
integrability is automatic in the full space
$\mathcal E$. On the embedded state--generator manifold, integrability
must be verified together with the tangency condition.

This distinction is essential for a stationary open system.  The
ambient space $\mathcal E=\mathcal R\oplus\mathcal H$ contains
arbitrary pairs $(\mathbf r,\mathbf h)$, but most such pairs do not
represent a physical stationary model.  The admissible pairs lie in
the constrained set
\[
\mathcal N
=
\left\{
(\mathbf r,\mathbf h)\in\mathcal E:
\mathcal L(\mathbf h)\rho(\mathbf r)=0,
\quad
\operatorname{Tr}\rho=1,
\quad
\rho\geq0
\right\},
\]
and the physical model manifold satisfies
\[
\Phi(\mathcal M)\subseteq\mathcal N\subset\mathcal E.
\]
$\Phi(\mathcal M)$ is selected by the null states of the
Liouvillian rather than by the ambient product geometry alone
\cite{Lindblad1976,Avron2000}.  When the stationary state is unique and
depends smoothly on the controls, differentiating its defining
equation gives
\[
\mathcal L\,\partial_\mu\rho_{\rm ss}
+(\partial_\mu\mathcal L)\rho_{\rm ss}=0,
\qquad
\operatorname{Tr}(\partial_\mu\rho_{\rm ss})=0.
\]
These relations constrain the allowed tangent vectors
$(\partial_\mu\mathbf r,\partial_\mu\mathbf h)$.  The canonical map
$J$ rotates arbitrary ambient state and generator variations into one
another, but it need not preserve this constrained tangent space.
K\"ahler compatibility is not automatic after restriction to
physical stationary states; it can hold only where the Liouvillian
null-state constraint happens to be invariant under the canonical
quarter-turn.

For a two-dimensional control manifold, this restriction can be tested
without explicitly constructing the induced $J$.  The Wirtinger bound
gives
\[
F_{g\omega}^{2}\leq\det G,
\]
and it is useful to define
\[
\kappa(g,\omega)
=\frac{F_{g\omega}^{2}}{\det G},
\qquad
0\leq\kappa\leq1.
\]
The equality $\kappa=1$ means that the canonical state--generator area
exhausts the Riemannian area of the physical tangent plane.  If
$\kappa<1$, part of the metric deformation of the stationary manifold
has no conjugate partner under the canonical two-form.

\begin{figure*}[t]
  \centering
  \includegraphics[width=0.485\textwidth]{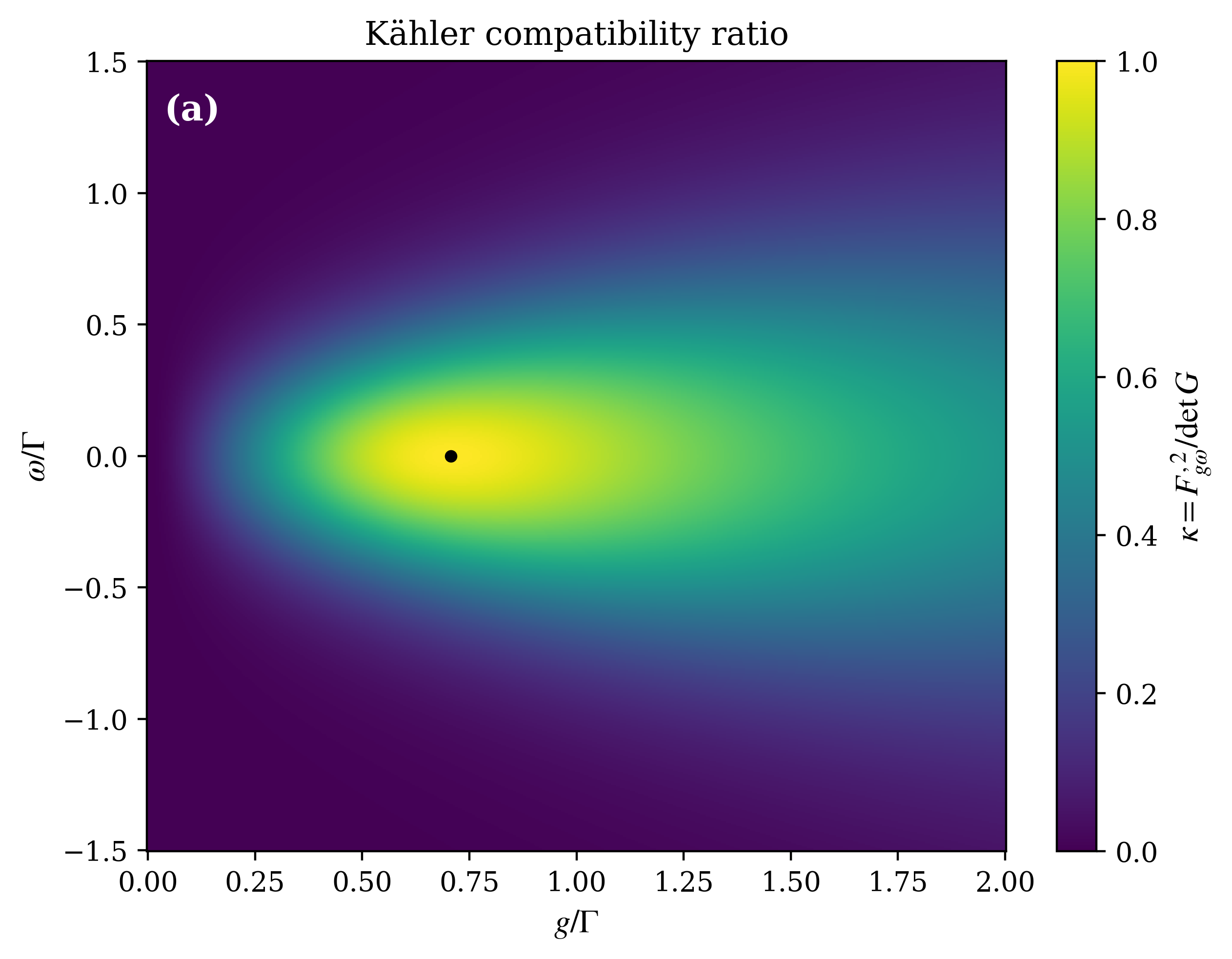}
  \hfill
  \includegraphics[width=0.485\textwidth]{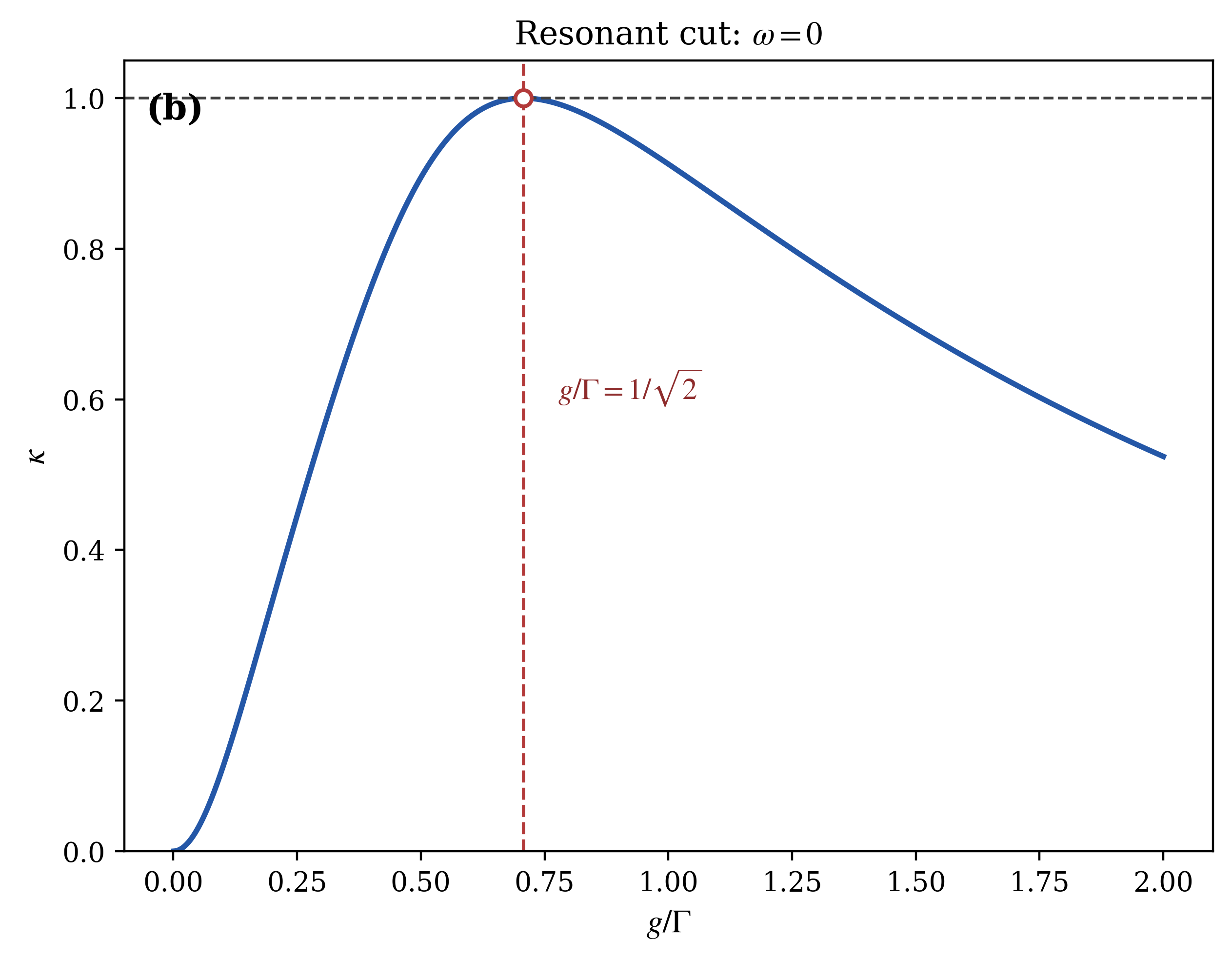}
  \caption{Metric--symplectic compatibility for the amplitude-damped
  optical Bloch model at fixed $\Gamma$. (a) The dimensionless ratio
  $\kappa=F_{g\omega}^{2}/\det G$ over the control plane.  The black
  marker denotes the compatible point. (b) Resonant section
  $\omega=0$, for which $\kappa$ reaches unity at
  $g/\Gamma=1/\sqrt{2}$.  At this point the response area form and
  Riemannian volume form agree in magnitude.}
  \label{fig:kahler-compatibility}
\end{figure*}

Figure~\ref{fig:kahler-compatibility} evaluates this criterion for the
amplitude-damped optical Bloch model developed explicitly in
Sec.~\ref{sec:examples}.  Panel (a) shows that the physical null-state
surface is generally not compatible: $\kappa$ remains strictly below
unity over nearly the entire $(g/\Gamma,\omega/\Gamma)$ plane.  The
single black point is not a numerical threshold or a broad compatible
phase.  It is an isolated saturation of the geometric bound.

The uniqueness is transparent on the resonant section.  With
\[
q=\frac{g}{\Gamma},
\qquad
\omega=0,
\]
the obstruction to compatibility factorizes exactly as
\[
\det G-F_{g\omega}^{2}
=
\frac{
(2q^2-1)^2
(4q^4+4q^2+5)
(4q^4+12q^2+1)
}{(2q^2+1)^6}.
\]
The last two factors are strictly positive for real $q$, and the first
factor vanishes only when
\[
2q^2=1.
\]
For the physical branch $g\geq0$, the unique compatible point is
\[
\boxed{
\omega=0,
\qquad
\frac{g}{\Gamma}=\frac{1}{\sqrt2}
}.
\]
The squared factor $(2q^2-1)^2$ also shows that the point is a
tangential saturation of the bound rather than a sign-changing
crossing.  Coherent rotation and amplitude damping are both finite at
this point.  Compatibility does not signal the absence of
dissipation; it identifies a precise matching between coherent
generator deformation and the dissipative contraction encoded in the
stationary-state response.

This interpretation should not be confused with Liouville's theorem.
The compatibility relation
\[
F(X,Y)=g(JX,Y)
\]
is a kinematic statement about tensors on the model manifold, whereas
preservation of a symplectic form or volume by a dynamical flow is a
separate dynamical statement.  Nevertheless, Fig.~\ref{fig:kahler-compatibility}
shows that the open-system null-state constraint controls whether the
metric and canonical response area can be mutually compatible.  The
isolated point is a central result: the physical stationary
submanifold is generically non-K\"ahler, yet finite coherent and
dissipative effects can balance at one model for which its tangent
geometry becomes exactly compatible.  This compatible tangent
structure supplies the geometric foundation for the covariant
differentiation and transport developed below.

\section{Levi--Civita Connection and Covariant Transport}
\label{sec:covariant-transport}

The practical motivation for a covariant transport law is the possibility of
reusing an observable calculation across control space.  Suppose, for example,
that a linear or nonlinear spectral response has been evaluated at a reference
control point $p$.  We would like to carry that result to a second point $q$
along a path in control space, rather than repeat the entire dynamical
calculation independently at every model.  Infinitesimally, this resembles a
response expansion in the control displacement.  Over a finite path, the
comparison is more subtle: the stationary state, generator, and physical
tangent frame all change along the path, so response objects at different
points do not naturally belong to the same tangent space.  This viewpoint
builds on our earlier geometric formulation of open-system thermodynamics, in
which quasistatic evolution follows a manifold of stationary states and work
acquired around a control-space cycle is expressed as a curvature flux
\cite{Bittner:2026aa}.  The present problem asks for the corresponding local
rule for carrying state and response information along such a manifold.

In our recent work on third-order multidimensional spectral response, we
demonstrated this strategy using a local expansion that is
essentially a Taylor transport in the control parameters \cite{JCP3}.  That
calculation showed how differential information obtained at a reference model
can be used to predict spectra at nearby points.  Ordinary derivatives remain
tied to the chosen control coordinates and implicitly identify
response objects living in different tangent spaces.  Here we develop the
geometric machinery required to make that identification intrinsic.  The
transport path is constrained to $\Phi(\mathcal M)$, the submanifold of
admissible stationary state--generator pairs selected by the Liouvillian
null-state condition, and its induced metric determines the Levi--Civita
connection.  Covariant differentiation and parallel transport then provide
the proper replacements for ordinary Taylor coefficients when carrying
states, perturbations, and ultimately observable-response tensors from one
control point to another.

\subsection{The Tangent Bundle and Induced Metric}

Section~\ref{sec:II} established the embedding $\Phi$ and the tangent
basis $\mathbf e_\mu=\partial_\mu\Phi$.  Its differential
\[
d\Phi:T_{\boldsymbol{\lambda}}\mathcal M
\longrightarrow
T_{\Phi(\boldsymbol{\lambda})}\mathcal E,
\]
identifies tangent vectors on model space with tangent vectors of its
physical image.  For a regular embedding, the Gram matrix
$g_{\mu\nu}=\mathbf e_\mu\cdot\mathbf e_\nu$ is invertible, with
\[
g^{\mu\rho}
g_{\rho\nu}
=
\delta^\mu_{\;\nu}.
\]
This inverse raises and lowers model-space indices,
\[
X_\mu
=
g_{\mu\nu}
X^\nu,
\qquad
X^\mu
=
g^{\mu\nu}
X_\nu.
\]

The new ingredient needed for transport is that this induced basis
varies over the manifold:
\[
\partial_\mu\mathbf e_\nu
\neq
0,
\]
in general. This variation produces the covariant structure developed below.
Differentiating a vector field
requires differentiating not only its components but also the basis
itself:
\[
\partial_\mu
\left(
X^\nu
\mathbf e_\nu
\right)
=
(\partial_\mu X^\nu)
\mathbf e_\nu
+
X^\nu
\partial_\mu\mathbf e_\nu.
\]

The second term records the changing tangent frame.  The
Levi--Civita connection removes the coordinate-dependent part of this
change and supplies the intrinsic derivative used below.

\subsection{Covariant Differentiation}

Coordinate derivatives of the state--generator embedding,
\[
\partial_\mu\Phi,\qquad
\partial_\mu\partial_\nu\Phi,\qquad\ldots,
\]
are not by themselves intrinsic geometric objects. Because the tangent basis varies across the embedded manifold, higher derivatives contain contributions from the changing basis as well as from variation of the physical response.

The Levi--Civita connection separates these effects. For a vector field
\[
X=X^\mu\mathbf e_\mu,
\]
the covariant derivative is
\[
\nabla_\mu X^\nu
=
\partial_\mu X^\nu
+
\Gamma^\nu_{\mu\rho}X^\rho.
\]
Successive covariant derivatives generate intrinsic higher-order response tensors,
\[
\nabla_\mu,\qquad
\nabla_\mu\nabla_\nu,\qquad
\nabla_\mu\nabla_\nu\nabla_\rho,\qquad\ldots,
\]
whose connection terms remove basis-dependent contributions at every order.
These objects transform covariantly under changes of coordinates on the
manifold of physical models and form the derivative hierarchy for geometric
response theory.

\subsection{Parallel Transport and Liouvillian Transport}

The Levi--Civita connection determines how vectors are transported
between neighboring points of the state--generator manifold. Let
\[
\gamma:s\mapsto\lambda^\mu(s)
\]
be a smooth curve in the manifold of physical models, and let
\[
V(s)=V^\mu(s)\mathbf e_\mu
\]
denote a vector field defined along this curve.

Two evolution equations associated with the connection play distinct roles.
An affinely parameterized geodesic transports its own tangent vector and obeys
\begin{equation}
\ddot\lambda^\alpha
+\Gamma^\alpha_{\mu\nu}[\lambda(s)]
\dot\lambda^\mu\dot\lambda^\nu=0.
\label{eq:geodesic-eom}
\end{equation}
This equation selects intrinsically straight paths through the manifold; with
fixed endpoints it defines the boundary-value problem used below.

The vector is said to undergo parallel transport if its covariant
derivative along the curve vanishes,
\[
\nabla_{\dot\gamma}V=0,
\]
where
\[
\dot\gamma
=
\frac{d\lambda^\mu}{ds}\,
\partial_\mu.
\]
In local coordinates this condition becomes
\begin{equation}
\dot V^\alpha
+\Gamma^\alpha_{\mu\nu}[\lambda(s)]
\dot\lambda^\mu V^\nu=0.
\label{eq:parallel-transport-eom}
\end{equation}
Unlike Eq.~\eqref{eq:geodesic-eom}, this equation transports an independent
tangent vector along a specified path; the path need not itself be a
geodesic.

Equation~\eqref{eq:parallel-transport-eom} defines the unique transport law
that preserves the
metric,
\[
\frac{d}{ds}
g(V,V)
=
0,
\]
so that lengths and inner products remain invariant under transport.

The Liouvillian transport operator introduced in the main text may be
viewed as the corresponding transport operator acting on observable
response objects rather than ordinary tangent vectors. Formally, the
transport equation may be written
\[
\nabla_\mu\mathcal O
=
\mathcal K_\mu
\mathcal O,
\]
where
\(
\mathcal K_\mu
\)
is the Liouvillian transport generator associated with the parameter
direction
\(
\lambda^\mu.
\)

Integrating this equation along a path
\(
\gamma
\)
yields the transport operator
\[
\mathsf T_\gamma
=
\mathcal P
\exp
\left[
\int_\gamma
d\lambda^\mu
\,
\mathcal K_\mu
\right],
\]
which transports observables between neighboring physical models.

The hierarchy is direct. The state--generator embedding induces the metric.
The metric determines the Levi--Civita connection. The connection
defines the covariant derivative, and the covariant derivative
generates the Liouvillian transport equation whose solution is the
path-ordered transport operator.

The resulting hierarchy is independent of the coordinate representation of
the parameter manifold. It establishes the intrinsic structure that any
explicit Liouvillian transport theory must preserve. The derivation of
local transport expansions and their application to nonlinear spectroscopic
pathways are taken up in the companion paper.

\section{Illustrative Examples}
\label{sec:examples}

\subsection{Amplitude-damped optical Bloch model}

The amplitude-damped optical Bloch model is a minimal example in
which the state--generator embedding, the response two-form, and the
Levi--Civita connection can all be constructed explicitly.  We take
\[
H(g,\omega)
=
\begin{pmatrix}
0 & g/2\\
g/2 & \omega
\end{pmatrix},
\qquad
C=\sqrt{\Gamma}\,\lvert 0\rangle\langle 1\rvert,
\]
where $g$ is the coherent coupling, $\omega$ is the detuning, and
$\Gamma$ is the amplitude-damping rate.  The density operator evolves
under
\[
\dot\rho
=-i[H,\rho]
+C\rho C^\dagger
-\frac{1}{2}\{C^\dagger C,\rho\}.
\]
For fixed $\Gamma$, the control manifold is coordinatized by
$\lambda=(g,\omega)$.  Solving $\mathcal L(\lambda)\rho_{\rm ss}=0$
together with $\operatorname{Tr}\rho_{\rm ss}=1$ gives the stationary
Bloch vector
\[
\mathbf r(g,\omega)
=
\frac{1}{D}
\begin{pmatrix}
-4g\omega\\
2\Gamma g\\
-(\Gamma^2+4\omega^2)
\end{pmatrix},
\qquad
D=\Gamma^2+2g^2+4\omega^2.
\]
We represent the Hamiltonian part of the generator by
\[
\mathbf h(g,\omega)=(g,0,\omega).
\]
The physical model defines the explicit embedding
\[
\Phi:(g,\omega)\longmapsto
\bigl(\mathbf r(g,\omega),\mathbf h(g,\omega)\bigr).
\]

The tangent vectors $\mathbf e_g=\partial_g\Phi$ and
$\mathbf e_\omega=\partial_\omega\Phi$ determine the induced metric
and response two-form without any additional geometric assumptions:
\begin{align}
G_{\mu\nu}
&=
\partial_\mu\mathbf r\cdot\partial_\nu\mathbf r
+\partial_\mu\mathbf h\cdot\partial_\nu\mathbf h,
\\
F_{\mu\nu}
&=
\partial_\mu\mathbf r\cdot\partial_\nu\mathbf h
-\partial_\nu\mathbf r\cdot\partial_\mu\mathbf h.
\end{align}
Evaluating these tensors over the control plane gives the compatibility
landscape in Fig.~\ref{fig:kahler-compatibility}.  Section~III shows
that its isolated saturation point occurs at $\omega=0$ and
$g/\Gamma=1/\sqrt2$; here we use the same explicit embedding to
construct and validate the Levi--Civita connection.

The same embedding fixes the Levi--Civita symbols,
\[
\Gamma^\alpha_{\mu\nu}
=\frac{1}{2}G^{\alpha\beta}
\left(
\partial_\mu G_{\beta\nu}
+\partial_\nu G_{\beta\mu}
-\partial_\beta G_{\mu\nu}
\right),
\qquad
\alpha,\mu,\nu\in\{g,\omega\}.
\]
Both geodesics and parallel transport are determined
directly by the stationary state and its generator through
Eqs.~\eqref{eq:geodesic-eom} and \eqref{eq:parallel-transport-eom}.
This example gives a closed, model-derived realization of
the full sequence
\[
\mathcal L(\lambda)
\longrightarrow
\rho_{\rm ss}(\lambda)
\longrightarrow
(\mathbf r,\mathbf h)
\longrightarrow
(G,F,\Gamma),
\]
which is the geometric input required for the Liouvillian-transport
theory developed in the companion work.  Symbolic and
finite-difference reference implementations used for this example are
included with the manuscript source.

\subsection{Automatic differentiation of the connection}

Automatic differentiation (AD) evaluates derivatives by systematic
application of the chain rule to the numerical program itself.  It differs
from symbolic differentiation and finite
differences: it introduces no differencing step, while retaining the
floating-point evaluation of the original model
\cite{GriewankWalther2008,Baydin2018}.  Differentiable-programming
approaches are now used broadly in molecular simulation and quantum
dynamics, including differentiable physics engines and master-equation
solvers \cite{SchoenholzCubuk2021,Craig2024DiffLindblad}.

Our implementation uses JAX, which supplies composable forward- and
reverse-mode transformations for NumPy-style numerical programs
\cite{jax2018github}.  The present control manifold is low dimensional,
so we use forward-mode Jacobian propagation (\texttt{jax.jacfwd}) for
both the state--generator map and the metric.  All calculations enable
64-bit arithmetic.  The archived calculation used JAX and jaxlib
version 0.11.0; package and environment details are recorded with the
source archive.

The Levi--Civita symbols need not be derived separately for every new
model.  If the stationary-state map and generator coordinates are
differentiable functions of the controls, automatic differentiation
yields the complete local geometry.  The sequence
\begin{align}
R_{i\mu}&=\partial_\mu r_i,
&
H_{a\mu}&=\partial_\mu h_a,
\\
G_{\mu\nu}&=(R^{\mathsf T}R+H^{\mathsf T}H)_{\mu\nu},
&
\partial_\eta G_{\mu\nu}
&=\operatorname{AD}_{\eta}[G_{\mu\nu}]
\end{align}
can be evaluated directly as a composition of Jacobians.  Contracting
the differentiated metric with $G^{-1}$ then gives
\[
\Gamma^\alpha_{\mu\nu}
=\frac{1}{2}G^{\alpha\beta}
\left(
\partial_\mu G_{\beta\nu}
+\partial_\nu G_{\beta\mu}
-\partial_\beta G_{\mu\nu}
\right).
\]
This procedure differentiates the geometry rather than imposing a
connection externally.

For a general open-system model, the stationary state can itself be
obtained from the constrained linear system
\[
\begin{pmatrix}
\widetilde{\mathcal L}(\lambda)\\
\langle\!\langle\mathbf 1\rvert
\end{pmatrix}
\lvert\rho_{\rm ss}(\lambda)\rangle\!\rangle
=
\begin{pmatrix}
0\\1
\end{pmatrix},
\]
where one dependent row of the Liouvillian is replaced by the trace
constraint.  Differentiating through this solve produces
$\partial_\mu\rho_{\rm ss}$ and hence $R_{i\mu}$ without requiring a
closed-form stationary state.  At the level of a generic constrained
system $M(\lambda)x(\lambda)=b(\lambda)$, the differentiated solution
obeys
\[
M\,\partial_\mu x
=
\partial_\mu b
-(\partial_\mu M)x,
\]
which JAX evaluates through its differentiable linear-algebra
primitives.  The procedure yields the
model-specific $G$, $F$, and $\Gamma$ from the Liouvillian alone,
provided the stationary state is isolated and the constrained system
is nonsingular.  Near degeneracies or rank-changing points, the
conditioning of this solve and of $G^{-1}$ must be monitored explicitly.
AD does not remove these mathematical singularities, and derivatives
of an ill-conditioned stationary state can be unreliable even when the
primal linear solve appears converged.

\subsection{Numerical validation of the differentiated geometry}

We validated the calculation by three independent routes.
The symbolic reference differentiates the closed-form Bloch vector with
SymPy and contracts the resulting exact expressions for $G$, $F$, and
$\Gamma$.  The automatic-differentiation route instead constructs the
$4\times4$ Liouvillian, replaces one dependent row by the trace
constraint, solves for $\rho_{\rm ss}$, and applies forward-mode
automatic differentiation through both the linear solve and the metric
calculation.  No closed-form expression for $\rho_{\rm ss}$ is used in
this second route.  Finally, an independent central-difference
calculation differentiates $\mathbf r$ and $\mathbf h$ to obtain $G$
and $F$, and then differentiates $G$ a second time to obtain the
connection.  The finite-difference displacement was
\[
\delta\lambda^mu
=10^{-4}\max(1,\lvert\lambda^mu\rvert).
\]

All calculations used dimensionless variables with $\Gamma=1$ and
64-bit arithmetic.  We tested an off-resonant weak-drive point
$(g,\omega)=(0.35,0.12)$, the compatible resonant point
$(1/\sqrt{2},0)$, and an off-resonant strong-drive point
$(1.20,-0.40)$.  For an object $A$, we report the componentwise maximum
absolute discrepancy from the symbolic result,
\[
\epsilon_\infty(A)
=\max_{i\ldots}\left|A^{\rm test}_{i\ldots}
-A^{\rm sym}_{i\ldots}\right|.
\]
The tensor components in this dimensionless test are of order unity,
so the absolute discrepancies also give the relevant scale of the
relative agreement.

Across all three points, differentiation through the Liouvillian solve
gave
\begin{align}
\epsilon_\infty(\mathbf r)&\leq 1.12\times10^{-16},
&
\epsilon_\infty(G)&\leq 8.89\times10^{-16},
\\
\epsilon_\infty(F)&\leq 4.45\times10^{-16},
&
\epsilon_\infty(\Gamma)&\leq 5.56\times10^{-16}.
\end{align}
Table~\ref{tab:bloch-geometry-validation} gives the connection errors
point by point.  The larger finite-difference discrepancy is expected:
the Christoffel symbols require a second numerical differentiation,
whereas automatic differentiation evaluates the required derivatives
without a step-size truncation error.

\begin{table}[t]
\caption{Validation of the optical Bloch Levi--Civita connection at
$\Gamma=1$.  Errors are maximum absolute componentwise differences
from the symbolic result.  The final column is the smallest eigenvalue
of the automatically differentiated metric.}
\label{tab:bloch-geometry-validation}
\begin{ruledtabular}
\begin{tabular}{cccc}
$(g,\omega)$ & $\epsilon_\infty(\Gamma_{\rm AD})$
& $\epsilon_\infty(\Gamma_{\rm FD})$ & $\lambda_{\min}(G)$\\
\hline
$(0.35,0.12)$ & $5.56\times10^{-16}$ & $1.33\times10^{-7}$ & $2.028$\\
$(1/\sqrt{2},0)$ & $1.67\times10^{-16}$ & $2.83\times10^{-8}$ & $1.500$\\
$(1.20,-0.40)$ & $2.22\times10^{-16}$ & $1.92\times10^{-8}$ & $1.122$\\
\end{tabular}
\end{ruledtabular}
\end{table}

At the compatible resonant point, the calculation gives the especially
simple tensors
\[
G=
\begin{pmatrix}
3/2&0\\0&3
\end{pmatrix},
\qquad
F=
\begin{pmatrix}
0&3/\sqrt{2}\\-3/\sqrt{2}&0
\end{pmatrix},
\]
for which $F_{g\omega}^2=\det G=9/2$.  With coordinate order
$(g,\omega)$, the only nonzero connection coefficients at this point
are
\[
\Gamma^{g}_{gg}=-\frac{\sqrt{2}}{3},
\qquad
\Gamma^{g}_{\omega\omega}=-\frac{2\sqrt{2}}{3}.
\]
The remaining coefficients vanish to numerical precision.  In
particular, the independently assembled connection is symmetric in its
lower indices at every test point,
\[
\max_{\alpha\mu\nu}
\left|\Gamma^\alpha_{\mu\nu}
-\Gamma^\alpha_{\nu\mu}\right|=0,
\]
as required for a torsion-free Levi--Civita connection.  The positive
minimum metric eigenvalues listed in
Table~\ref{tab:bloch-geometry-validation} also verify that the inverse
metric used in the contraction is well defined at all three points.

These tests establish more than agreement with a single closed-form
formula.  They validate the full computational chain from the
Liouvillian and normalization constraint through the stationary state,
the state--generator Jacobian, the induced metric, and its connection.
For a new model with an isolated differentiable stationary state,
the Levi--Civita symbols can be generated directly from the model
Liouvillian without rederiving model-specific expressions by hand.
This focused use of automatic differentiation is part of the
state--generator geometry; path-ordered Liouvillian transport and its
numerical implementation remain the subject of the companion paper.

\subsection{A geodesic between two control points}

As a first transport calculation, consider the fixed-$\Gamma$ control
manifold and connect the reference model
\[
p:(\omega/\Gamma,g/\Gamma)=(0,0)
\]
to the known target
\[
q:(\omega/\Gamma,g/\Gamma)=(0.40,1.00).
\]
The path is not prescribed as a linear interpolation of the control
parameters.  We solve Eq.~\eqref{eq:geodesic-eom} as the two-point
boundary-value problem
\[
\lambda(0)=p,
\qquad
\lambda(1)=q,
\]
where the connection at every collocation point is evaluated from the
symbolic optical Bloch metric.  A straight control-space segment was
used only as the initial guess for the boundary-value solver.  The
converged solution had a maximum collocation rms residual of
$9.97\times10^{-9}$.

The length of a candidate path was evaluated with the physical metric,
\[
L[\lambda]
=\int_0^1
\sqrt{
\dot\lambda^\mu G_{\mu\nu}[\lambda(s)]
\dot\lambda^\nu
}\,ds.
\]
The Levi--Civita geodesic has length
\[
L_{\rm geo}=1.53947,
\]
whereas the straight interpolation between the same endpoints has
\[
L_{\rm lin}=1.57784.
\]
The model-derived geodesic reduces the physical path length by
$0.03836$, or approximately $2.43\%$, even for this modest displacement
in control space.

\begin{figure*}[t]
  \centering
  \includegraphics[width=0.94\textwidth]{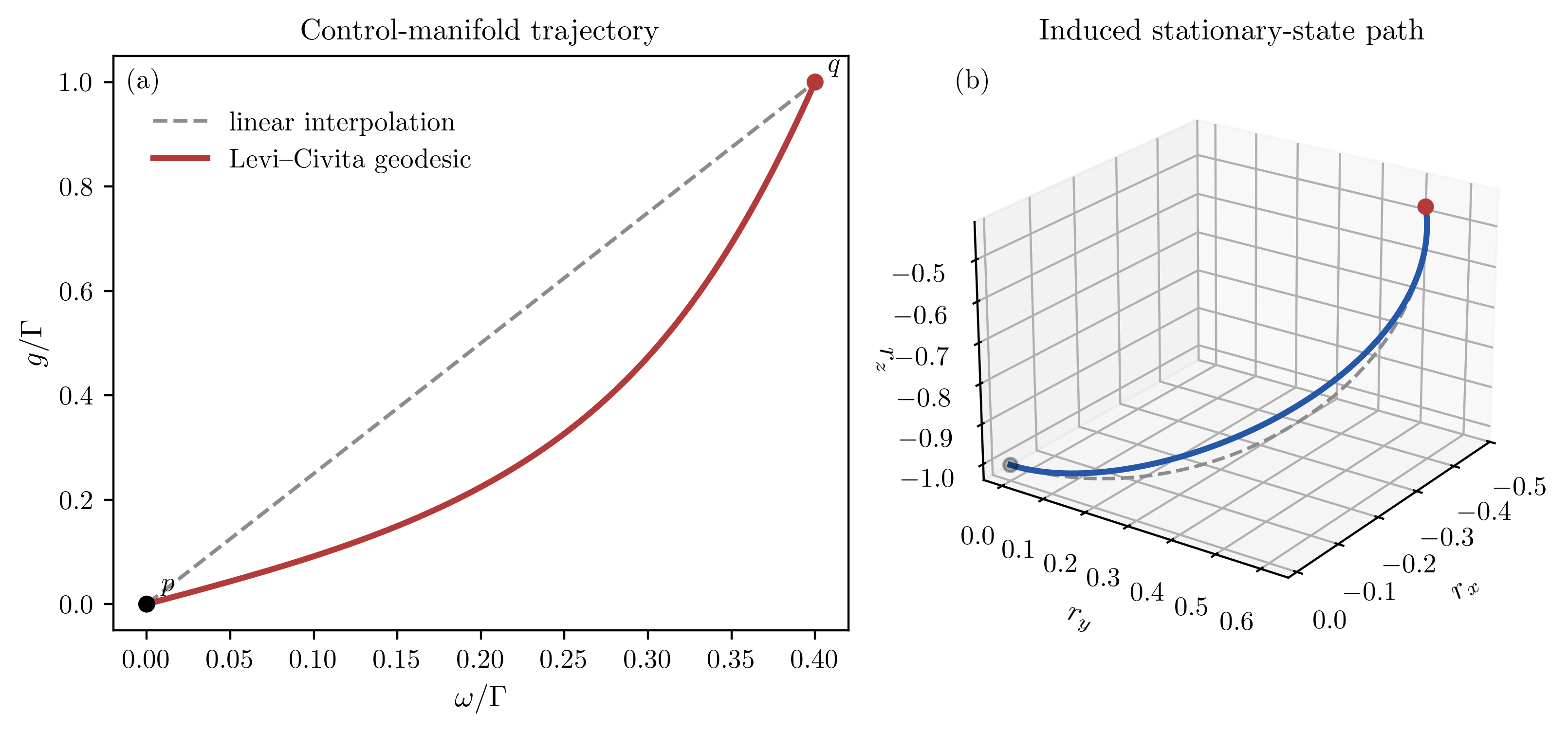}
  \caption{Geodesic transport between two optical Bloch control points
  at fixed $\Gamma$. (a) The Levi--Civita geodesic (red) from
  $p=(0,0)$ to $q=(0.40,1.00)$ in $(\omega/\Gamma,g/\Gamma)$ coordinates,
  compared with linear parameter interpolation (gray dashed).  The
  geodesic bends toward smaller $g$ because distance is measured by the
  induced state--generator metric rather than by the Euclidean control
  coordinates. (b) Images of the same two control-space paths under the
  stationary-state map $\lambda\mapsto\mathbf r(\lambda)$.  The blue
  curve is the state trajectory induced by the geodesic; the dashed
  curve is induced by straight interpolation.}
  \label{fig:bloch-geodesic}
\end{figure*}

Figure~\ref{fig:bloch-geodesic}(a) shows the resulting trajectory in
the control manifold.  Its curvature is a direct consequence of the
spatially varying $G_{\mu\nu}$ and $\Gamma^\alpha_{\mu\nu}$; it is not
introduced by an external cost function.  Figure~\ref{fig:bloch-geodesic}(b)
shows the corresponding trajectory of stationary states in the Bloch
representation.  Although the two paths share identical physical
endpoints, their intermediate stationary states differ substantially.
This calculation determines the intrinsic path along which a state,
tangent vector, or response object should subsequently be parallel
transported.  Computing that object requires solving its covariant evolution
equation along the geodesic and is the next step.

\section{Discussion}
\label{sec:discussion}

\subsection{The physical geometry of stationary models}

The state--generator pair is the minimal object needed to describe the
geometry of a stationary dynamical model.  The state identifies the realized
solution; the generator identifies the dynamics that makes it stationary.
Taken together, they define an embedding
$\Phi:\mathcal M\rightarrow\mathcal R\oplus\mathcal H$ whose differential
contains both pieces of response.  The metric $g$, response potential $A$,
and closed two-form $F$ are different contractions of this same differential.
The symmetric and antisymmetric sectors are not separate additions to control
space.  They are two views of how the full physical model changes.

This result extends our earlier curvature description of geometric work in
open quantum systems \cite{Bittner:2026aa,JCP2}.  That work identified the
antisymmetric response with an oriented area in control space.  The present
theory identifies the space in which that area lives and supplies its metric.
It also exposes a constraint that is easy to miss when state and generator
variations are treated separately.  The ambient space
$\mathcal E=\mathcal R\oplus\mathcal H$ contains arbitrary pairs, while the
physical image $\Phi(\mathcal M)$ contains only those selected by the
Liouvillian null-state equation.  Its tangent vectors must satisfy the
differentiated stationarity condition.  The geometry of stationary response
is the geometry of this constrained surface, not of the ambient product
space.

This distinction resolves the compatibility question raised in our previous
work.  The canonical quarter-turn $J$ is compatible with the Euclidean metric
and symplectic form in the ambient space, but it does not generally preserve
the tangent plane of the physical surface.  The optical Bloch model shows the
obstruction directly.  Its compatibility ratio
$\kappa=F_{g\omega}^{2}/\det G$ remains below unity across almost the entire
control plane and reaches unity only at
$\omega=0$, $g/\Gamma=1/\sqrt{2}$.  Exact factorization shows that this point
is an isolated tangential saturation, not the edge of a compatible phase.

Both coherent driving and dissipative relaxation remain finite at the
compatible point.  Compatibility does not mark a return to closed-system
dynamics.  It marks a local balance: state and generator variations span a
physical tangent plane invariant under the canonical state--generator
rotation.  Away from this point, part of the metric deformation has no
conjugate partner in the response area form.  The failure of K\"ahler
compatibility is not imposed by the ambient geometry.  It enters through the
dynamical constraint linking a stationary state to the Liouvillian that
annihilates it.

\subsection{From local response to transport}

K\"ahler compatibility and covariant transport answer different questions.
Compatibility asks whether $g$, $F$, and $J$ close into a common tangent
structure.  Transport asks how to compare tangent or response objects at
different points.  The second problem requires only a regular induced metric.
Wherever $\Phi$ is regular, $g$ defines a unique Levi--Civita connection,
including regions where $\kappa<1$.  The isolated compatible point is special,
but it is not the only point at which geodesics, covariant derivatives, or
parallel transport exist.

The geodesic calculation gives this distinction a physical meaning.  A
straight segment in $(\omega/\Gamma,g/\Gamma)$ interpolates control values.  A
geodesic minimizes distance between the corresponding state--generator
models.  The two paths differ because equal coordinate displacements need not
produce equal physical changes.  Even for the two-level model, the geodesic
bends away from the straight control-space path and passes through a different
sequence of stationary states.  In larger control spaces, where response can
be strongly anisotropic, this difference should become more pronounced.

The transport problem also sharpens the relation to spectroscopy.  In our
treatment of third-order multidimensional response, a Taylor-like expansion
at one control point predicted spectra at nearby points \cite{JCP3}.  Such an
expansion works locally, but ordinary derivatives continue to use the frame at
the reference model.  Over a finite displacement, the frame changes along
with the state and generator.  Covariant derivatives subtract this frame
motion order by order.  Parallel transport supplies the finite-path
comparison, and path ordering records the noncommuting sequence of local
transport generators.

This geometric transport does not replace the underlying response
calculation.  It asks a different, practical question: once an observable has
been calculated at one model, what information is needed to carry it to
another?  The answer consists of three parts: the admissible stationary
manifold, a path on that manifold, and a connection that compares neighboring
tangent spaces.  The present paper supplies these ingredients.  The
Liouvillian transport operator determines how a particular spectral object
moves through them.

\subsection{Model comparison and non-Markovian extensions}

The induced geometry can also compare dynamical models, even when transport
of a particular observable is not the immediate goal.  Two open systems may
relax to the same stationary density operator while differing in decay rates,
coherent couplings, or dissipative channels.  A state-only description places
them at the same point.  The state--generator representation separates them.
The metric measures their local dissimilarity, geodesics identify economical
interpolations between them, and curvature or compatibility distinguishes
families that cannot be related by a flat change of control coordinates.  The
rank of $d\Phi$ also detects redundant controls: directions that change the
parameterization without changing the represented model appear as null
directions before restriction to a regular manifold.

These comparisons require a shared representation.  Operator bases must be
related by norm-preserving transformations, and state and generator
coordinates must use the same relative normalization.  Subject to those
conventions, scalar quantities such as geodesic distance, curvature
invariants, and the two-dimensional compatibility ratio do not depend on the
chosen control coordinates.  They offer a practical set of geometric labels
for comparing models, locating dynamically distinct regimes, and selecting
reference points for response transport.  They do not amount to a complete
classification: inequivalent models can share the same finite collection of
scalar invariants.

The curve $\gamma:s\mapsto\boldsymbol\lambda(s)$ used for geodesics and
parallel transport is a curve in model space, not a temporal trajectory of the
reduced state.  If the controls are changed slowly in an experiment, the
construction describes the quasistationary limit in which the system relaxes
at each point.  No temporal history is carried from one stationary model to
the next.  Parallel transport can still depend on the chosen control-space
path when the connection has curvature.  Such holonomy is geometric path
dependence, not environmental memory.

The Markovian restriction belongs to the model descriptor used in this paper.
We represent each control point by a time-homogeneous Liouvillian and its null
state, $\mathcal L\rho_{\rm ss}=0$.  A non-Markovian family could enter the
same parametric construction if every control point supplies a unique smooth
stationary state and a consistent representation of its dynamics.  For a
Nakajima--Zwanzig theory, that descriptor may be a memory kernel rather than a
finite instantaneous generator \cite{Nakajima1958,Zwanzig1960,Breuer2002}.
Alternatively, auxiliary-mode or hierarchical formulations may provide a
Markovian generator on an enlarged state space
\cite{Tanimura1989,Ishizaki2009}.  In either case, memory is a property of the
model assigned to each point, not of motion through parameter space.  The
choice of kernel representation or enlarged embedding also fixes the ambient
inner product and can change the induced metric; comparing those choices lies
beyond the present work.

\subsection{Computational reach and open questions}

Automatic differentiation moves this geometry beyond the few models that can
be solved on paper.  Starting from a differentiable Liouvillian, one can solve
the constrained stationary-state problem and differentiate through that solve
to obtain the state Jacobian, $G$, $F$, and the Levi--Civita symbols.  For the
optical Bloch model, the differentiated connection agrees with the symbolic
result to nearly machine precision.  Nested finite differences do not reach
the same accuracy because the connection requires another derivative of the
metric.  The relevant gain is not simply numerical precision: a new model no
longer requires a hand derivation of every Christoffel symbol.

Automatic differentiation does not regularize the physics.  A closing
Liouvillian gap, a nonunique stationary state, an ill-conditioned trace
constraint, or a singular induced metric remains singular after
differentiation.  These failures carry information.  They can signal a change
in the dimension of the stationary subspace, a dissipative transition, an
exceptional point, or a loss of rank in the chosen controls.  Near such
points, the smooth-manifold description used here must give way to a
stratified or bundle-valued treatment.  Positivity boundaries may require a
similar extension.

The metric also requires a declared normalization of state and generator
coordinates, including their relative scale.  This choice does not spoil
coordinate covariance, but it fixes the physical meaning of distance.
Comparisons between models must use the same convention.  Higher-dimensional
control spaces bring another complication: the two-dimensional identity
$F_{g\omega}^{2}=\det G$ no longer suffices.  Tangency of $J$ must be tested on
the full physical tangent bundle, followed by integrability if a K\"ahler
structure is claimed.

Several questions now become concrete.  How does the compatibility locus
change near a closing Liouvillian gap?  Can curvature or holonomy identify
regions where spectral transport becomes strongly path dependent?  Is the
shortest state--generator path also the most accurate or economical path for
transporting an observable?  And can the connection be reconstructed from
measured cross-response data without full state tomography?  These questions
link stationary-state geometry to the Liouvillian transport of nonlinear
spectral pathways.  The present theory fixes the manifold, metric, and
connection on which that program can be carried out.

\section*{Data Availability Statement}
The numerical data that support the findings of this study are openly
available in the Borealis Dataverse Repository at
\url{http://doi.org/[doi]}, reference number [reference number will be added
before publication].  The Python code and figure-generation scripts are
archived in Zenodo at \url{https://doi.org/10.5281/zenodo.22012960}.

\begin{acknowledgments}
CSA acknowledges funding from the Government of Canada (Canada Excellence
Research Chair CERC-2022-00055), from the Institut Courtois, Facult\'e des arts
et des sciences, Universit\'e de Montr\'eal (Chaire de recherche de direction de
l'Institut Courtois), and from the Natural Science and Engineering Research
Council of Canada (NSERC Discovery Grant RGPIN-2024-05893). ERB acknowledges
funding from the National Science Foundation (CHE-2404788), the Robert A.\ Welch
Foundation (E-1337), and by the U.S. Department of Energy, Office of Science,
under Award No. DE-SC0025706. ERB gratefully acknowledges funding from IVADO
for a Visiting Professorship at the Institut Courtois, Universit\'e de
Montr\'eal.
\end{acknowledgments}

\subsection*{Use of Generative Artificial Intelligence}
In compliance with institutional guidelines of the Universit\'e de Montr\'eal,
generative artificial intelligence tools were used to assist with the editing
of language and stylistic refinement of parts of the manuscript and to assist
in the synthesis of the literature. These tools were not used to generate
scientific content, perform analysis, or influence the interpretation of
results. All content has been reviewed and validated by the authors, who
assume full responsibility for the manuscript.

\appendix
\section{Notation and Conventions}
\label{app:notation}

This appendix collects the notation used throughout the manuscript and
fixes conventions whose signs or index order matter in the geometry.

\subsection{Spaces, coordinates, and indices}

The symbol $\mathcal M$ denotes the smooth manifold of admissible
physical models.  A point of $\mathcal M$ is specified locally by the
control coordinates
\[
\boldsymbol\lambda
=(\lambda^1,\ldots,\lambda^n).
\]
The ambient state--generator space is the real direct sum
\[
\mathcal E=\mathcal R\oplus\mathcal H,
\]
where $\mathcal R$ is the real coordinate space used to represent the
state and $\mathcal H$ is the corresponding real coordinate space for
the generator.  The state--generator map is
\[
\Phi:\mathcal M\longrightarrow\mathcal E,
\qquad
\Phi(\boldsymbol\lambda)
=\bigl(\mathbf r(\boldsymbol\lambda),
       \mathbf h(\boldsymbol\lambda)\bigr).
\]
The physical model manifold is the image $\Phi(\mathcal M)$; it need
not fill the ambient space.

Greek indices $\mu,\nu,\rho,\ldots$ label control coordinates on
$\mathcal M$.  Repeated upper and lower Greek indices are summed unless
an explicit summation sign is shown.  Latin component indices
$i,j,\ldots$ may be used for state coordinates and $a,b,\ldots$ for
generator coordinates.  Bold symbols denote coordinate vectors in the
ambient spaces.  The coordinate derivative and tangent basis are
\[
\partial_\mu
\equiv\frac{\partial}{\partial\lambda^\mu},
\qquad
\mathbf e_\mu
=\partial_\mu\Phi
=\bigl(\partial_\mu\mathbf r,
       \partial_\mu\mathbf h\bigr).
\]
A tangent vector is written equivalently as
\[
X=X^\mu\mathbf e_\mu=(X_r,X_h),
\]
with
\[
X_r=X^\mu\partial_\mu\mathbf r,
\qquad
X_h=X^\mu\partial_\mu\mathbf h.
\]
The centered dot denotes the Euclidean inner product in the chosen real
coordinate representation.  When the coordinates are coefficients in
an orthonormal operator basis, this is the corresponding
Hilbert--Schmidt inner product.

\subsection{Metric, forms, and complex structure}

The ambient product metric is pulled back to $\mathcal M$.  Its
components are
\[
g_{\mu\nu}
=\partial_\mu\mathbf r\cdot\partial_\nu\mathbf r
+\partial_\mu\mathbf h\cdot\partial_\nu\mathbf h,
\qquad
ds^2=g_{\mu\nu},d\lambda^\mu d\lambda^\nu.
\]
The inverse metric satisfies
\[
g^{\mu\rho}g_{\rho\nu}=\delta^\mu{}_{\nu}.
\]
In the optical Bloch example, $G_{\mu\nu}$ denotes the numerical matrix
representing this same induced metric; $G$ is not a second metric.
This capitalization keeps the model-specific matrix distinct from the
scalar coupling $g$ used in that example.

The response one-form is
\[
A=\mathbf r\cdot d\mathbf h
=A_\mu d\lambda^\mu,
\qquad
A_\mu=\mathbf r\cdot\partial_\mu\mathbf h.
\]
The orientation convention adopted throughout is
\[
F=dA=d\mathbf r\wedge d\mathbf h.
\]
In components,
\[
F
=\frac12 F_{\mu\nu},
d\lambda^\mu\wedge d\lambda^\nu,
\]
with
\[
F_{\mu\nu}
=\partial_\mu\mathbf r\cdot\partial_\nu\mathbf h
-\partial_\nu\mathbf r\cdot\partial_\mu\mathbf h,
\qquad
F_{\mu\nu}=-F_{\nu\mu}.
\]
This fixes the sign of every plotted $F_{g\omega}$.  Reversing the wedge
order reverses $F$ but leaves quadratic quantities such as
$F_{g\omega}^2/\det G$ unchanged.

The canonical ambient almost-complex structure is
\[
J(X_r,X_h)=(-X_h,X_r),
\qquad
J=
\begin{pmatrix}
0&-I\\ I&0
\end{pmatrix}.
\]
With the orientation above,
\[
J^2=-I,
\qquad
g(JX,JY)=g(X,Y),
\qquad
F(X,Y)=g(JX,Y).
\]
The Hermitian combination is denoted
\[
\mathscr H=g+iF.
\]
The script letter distinguishes this tensor from the generator
coordinate space $\mathcal H$.  The exterior derivative is denoted by
$d$; the identity $F=dA$ gives $dF=0$.

\subsection{Connection, curves, and transport}

The Christoffel symbols of the induced Levi--Civita connection use the
index order
\[
\Gamma^\alpha_{\mu\nu}
=\frac12 g^{\alpha\beta}
\left(
\partial_\mu g_{\beta\nu}
+\partial_\nu g_{\beta\mu}
-\partial_\beta g_{\mu\nu}
\right).
\]
The first index is the output component; the two lower indices are the
derivative direction and vector component, respectively.  The
connection is torsion free,
\[
\Gamma^\alpha_{\mu\nu}
=\Gamma^\alpha_{\nu\mu},
\]
and metric compatible, $\nabla_\alpha g_{\mu\nu}=0$.  For
$X=X^\mu\mathbf e_\mu$,
\[
\nabla_\mu X^\nu
=\partial_\mu X^\nu
+\Gamma^\nu_{\mu\rho}X^\rho.
\]

A curve is written
\[
\gamma:s\longmapsto\lambda^\mu(s),
\]
and an overdot denotes differentiation with respect to its affine
parameter $s$.  The geodesic and parallel-transport equations are given in
Eqs.~\eqref{eq:geodesic-eom} and \eqref{eq:parallel-transport-eom},
respectively.
The symbols $p$ and $q$ denote the initial and final points of a curve.
The path symbol $\gamma$ should not be confused with the optical Bloch
damping rate $\Gamma$ or the indexed connection
$\Gamma^\alpha_{\mu\nu}$.

\subsection{Operators and Liouville-space notation}

Operators on the Hilbert space are denoted by roman or Greek letters,
with $H$ the Hamiltonian, $C$ a collapse operator, and $\rho$ a density
operator.  The dagger denotes the Hilbert-space adjoint and
$\operatorname{Tr}$ the trace.  The dynamical generator is the
Liouvillian superoperator $\mathcal L(\boldsymbol\lambda)$, and a
stationary state satisfies
\[
\mathcal L(\boldsymbol\lambda)\rho_{\rm ss}=0,
\qquad
\operatorname{Tr}\rho_{\rm ss}=1.
\]
Double kets denote a fixed vectorization of operators,
\[
\lvert\rho\rangle\!\rangle,
\]
and the identity trace covector is
$\langle\!\langle\mathbf1\rvert$, so that
\[
\langle\!\langle\mathbf1\vert\rho\rangle\!\rangle
=\operatorname{Tr}\rho.
\]
The tilde in $\widetilde{\mathcal L}$ denotes the constrained linear
system obtained by replacing one dependent Liouvillian row with this
trace condition.  It does not denote a second physical generator.

\subsection{Optical Bloch example and numerical conventions}

For the example in Sec.~\ref{sec:examples}, $g$ is the coherent
coupling, $\omega$ the detuning, and $\Gamma$ the amplitude-damping
rate.  These symbols are scalars and are distinguished by context from
the metric $g_{\mu\nu}$ and Christoffel symbols
$\Gamma^\alpha_{\mu\nu}$.  Plots use the dimensionless controls
\[
\left(\frac{\omega}{\Gamma},\frac{g}{\Gamma}\right),
\]
with $\Gamma=1$ in the numerical calculations.  The source code stores
control arrays internally in the order $(g,\omega)$ but displays and
reports endpoints in the order $(\omega,g)$.

The abbreviation AD means automatic differentiation, while FD means
central finite differentiation.  Reported tensor errors use the
componentwise maximum absolute norm
\[
\epsilon_\infty(A)
=\max_{i\ldots}
\left|A^{\rm test}_{i\ldots}-A^{\rm sym}_{i\ldots}\right|.
\]
Unless otherwise stated, numerical quantities in the example are
dimensionless and automatic-differentiation calculations use 64-bit
arithmetic.

\section{Supporting Derivations}
\label{app:derivations}

\bibliography{references,Refs-Local-clean,auto-diff-refs}

\begin{thebibliography}{22}
\makeatletter
\providecommand \@ifxundefined [1]{
 \@ifx{#1\undefined}
}
\providecommand \@ifnum [1]{
 \ifnum #1\expandafter \@firstoftwo
 \else \expandafter \@secondoftwo
 \fi
}
\providecommand \@ifx [1]{
 \ifx #1\expandafter \@firstoftwo
 \else \expandafter \@secondoftwo
 \fi
}
\providecommand \natexlab [1]{#1}
\providecommand \enquote  [1]{``#1''}
\providecommand \bibnamefont  [1]{#1}
\providecommand \bibfnamefont [1]{#1}
\providecommand \citenamefont [1]{#1}
\providecommand \href@noop [0]{\@secondoftwo}
\providecommand \href [0]{\begingroup \@sanitize@url \@href}
\providecommand \@href[1]{\@@startlink{#1}\@@href}
\providecommand \@@href[1]{\endgroup#1\@@endlink}
\providecommand \@sanitize@url [0]{\catcode `\\12\catcode `\$12\catcode
  `\&12\catcode `\#12\catcode `\^12\catcode `\_12\catcode `\%12\relax}
\providecommand \@@startlink[1]{}
\providecommand \@@endlink[0]{}
\providecommand \url  [0]{\begingroup\@sanitize@url \@url }
\providecommand \@url [1]{\endgroup\@href {#1}{\urlprefix }}
\providecommand \urlprefix  [0]{URL }
\providecommand \Eprint [0]{\href }
\providecommand \doibase [0]{https://doi.org/}
\providecommand \selectlanguage [0]{\@gobble}
\providecommand \bibinfo  [0]{\@secondoftwo}
\providecommand \bibfield  [0]{\@secondoftwo}
\providecommand \translation [1]{[#1]}
\providecommand \BibitemOpen [0]{}
\providecommand \bibitemStop [0]{}
\providecommand \bibitemNoStop [0]{.\EOS\space}
\providecommand \EOS [0]{\spacefactor3000\relax}
\providecommand \BibitemShut  [1]{\csname bibitem#1\endcsname}
\let\auto@bib@innerbib\@empty

\bibitem [{\citenamefont {Weinhold}(1975)}]{Weinhold1975}
  \BibitemOpen
  \bibfield  {author} {\bibinfo {author} {\bibfnamefont {F.}~\bibnamefont
  {Weinhold}},\ }\bibfield  {title} {\enquote {\bibinfo {title} {Metric
  geometry of equilibrium thermodynamics},}\ }\href
  {https://doi.org/10.1063/1.431689} {\bibfield  {journal} {\bibinfo  {journal}
  {J. Chem. Phys.}\ }\textbf {\bibinfo {volume} {63}},\ \bibinfo {pages}
  {2479--2483} (\bibinfo {year} {1975})}\BibitemShut {NoStop}
\bibitem [{\citenamefont {Ruppeiner}(1995)}]{Ruppeiner1995}
  \BibitemOpen
  \bibfield  {author} {\bibinfo {author} {\bibfnamefont {G.}~\bibnamefont
  {Ruppeiner}},\ }\bibfield  {title} {\enquote {\bibinfo {title} {Riemannian
  geometry in thermodynamic fluctuation theory},}\ }\href
  {https://doi.org/10.1103/RevModPhys.67.605} {\bibfield  {journal} {\bibinfo
  {journal} {Rev. Mod. Phys.}\ }\textbf {\bibinfo {volume} {67}},\ \bibinfo
  {pages} {605--659} (\bibinfo {year} {1995})}\BibitemShut {NoStop}
\bibitem [{\citenamefont {Berry}(1984)}]{Berry1984}
  \BibitemOpen
  \bibfield  {author} {\bibinfo {author} {\bibfnamefont {M.~V.}\ \bibnamefont
  {Berry}},\ }\bibfield  {title} {\enquote {\bibinfo {title} {Quantal phase
  factors accompanying adiabatic changes},}\ }\href
  {https://doi.org/10.1098/rspa.1984.0023} {\bibfield  {journal} {\bibinfo
  {journal} {Proc. R. Soc. Lond. A}\ }\textbf {\bibinfo {volume} {392}},\
  \bibinfo {pages} {45--57} (\bibinfo {year} {1984})}\BibitemShut {NoStop}
\bibitem [{\citenamefont {Provost}\ and\ \citenamefont
  {Vallee}(1980)}]{Provost1980}
  \BibitemOpen
  \bibfield  {author} {\bibinfo {author} {\bibfnamefont {J.~P.}\ \bibnamefont
  {Provost}}\ and\ \bibinfo {author} {\bibfnamefont {G.}~\bibnamefont
  {Vallee}},\ }\bibfield  {title} {\enquote {\bibinfo {title} {Riemannian
  structure on manifolds of quantum states},}\ }\href
  {https://doi.org/10.1007/BF02193559} {\bibfield  {journal} {\bibinfo
  {journal} {Commun. Math. Phys.}\ }\textbf {\bibinfo {volume} {76}},\ \bibinfo
  {pages} {289--301} (\bibinfo {year} {1980})}\BibitemShut {NoStop}
\bibitem [{\citenamefont {Zanardi}, \citenamefont {Giorda},\ and\ \citenamefont
  {Cozzini}(2007)}]{Zanardi2007}
  \BibitemOpen
  \bibfield  {author} {\bibinfo {author} {\bibfnamefont {P.}~\bibnamefont
  {Zanardi}}, \bibinfo {author} {\bibfnamefont {P.}~\bibnamefont {Giorda}},\
  and\ \bibinfo {author} {\bibfnamefont {M.}~\bibnamefont {Cozzini}},\
  }\bibfield  {title} {\enquote {\bibinfo {title} {Information-theoretic
  differential geometry of quantum phase transitions},}\ }\href
  {https://doi.org/10.1103/PhysRevLett.99.100603} {\bibfield  {journal}
  {\bibinfo  {journal} {Phys. Rev. Lett.}\ }\textbf {\bibinfo {volume} {99}},\
  \bibinfo {pages} {100603} (\bibinfo {year} {2007})}\BibitemShut {NoStop}
\bibitem [{\citenamefont {Gu}(2010)}]{Gu2010}
  \BibitemOpen
  \bibfield  {author} {\bibinfo {author} {\bibfnamefont {S.-J.}\ \bibnamefont
  {Gu}},\ }\bibfield  {title} {\enquote {\bibinfo {title} {Fidelity approach to
  quantum phase transitions},}\ }\href@noop {} {\bibfield  {journal} {\bibinfo
  {journal} {Int. J. Mod. Phys. B}\ }\textbf {\bibinfo {volume} {24}},\
  \bibinfo {pages} {4371--4458} (\bibinfo {year} {2010})}\BibitemShut {NoStop}
\bibitem [{\citenamefont {Bittner}(2026)}]{Bittner:2026aa}
  \BibitemOpen
  \bibfield  {author} {\bibinfo {author} {\bibfnamefont {E.~R.}\ \bibnamefont
  {Bittner}},\ }\bibfield  {title} {\enquote {\bibinfo {title} {Geometric
  thermodynamics in open quantum systems: Coherence, curvature, and work},}\
  }\href {https://doi.org/10.1063/5.0335651} {\bibfield  {journal} {\bibinfo
  {journal} {The Journal of Chemical Physics}\ }\textbf {\bibinfo {volume}
  {164}},\ \bibinfo {pages} {194116} (\bibinfo {year} {2026})}\BibitemShut
  {NoStop}
\bibitem [{\citenamefont {Bittner}\ and\ \citenamefont
  {Silva-Acuña}(2026)}]{JCP2}
  \BibitemOpen
  \bibfield  {author} {\bibinfo {author} {\bibfnamefont {E.~R.}\ \bibnamefont
  {Bittner}}\ and\ \bibinfo {author} {\bibfnamefont {C.}~\bibnamefont
  {Silva-Acuña}},\ }\href {https://arxiv.org/abs/2606.22517} {\enquote
  {\bibinfo {title} {Geometric response in open quantum systems: Coherence,
  curvature, and susceptibility},}\ } (\bibinfo {year} {2026}),\ \Eprint
  {https://arxiv.org/abs/2606.22517} {arXiv:2606.22517 [quant-ph]} \BibitemShut
  {NoStop}
\bibitem [{\citenamefont {Gorini}, \citenamefont {Kossakowski},\ and\
  \citenamefont {Sudarshan}(1976)}]{Gorini1976}
  \BibitemOpen
  \bibfield  {author} {\bibinfo {author} {\bibfnamefont {V.}~\bibnamefont
  {Gorini}}, \bibinfo {author} {\bibfnamefont {A.}~\bibnamefont
  {Kossakowski}},\ and\ \bibinfo {author} {\bibfnamefont {E.~C.~G.}\
  \bibnamefont {Sudarshan}},\ }\bibfield  {title} {\enquote {\bibinfo {title}
  {Completely positive dynamical semigroups of {$N$}-level systems},}\ }\href
  {https://doi.org/10.1063/1.522979} {\bibfield  {journal} {\bibinfo  {journal}
  {J. Math. Phys.}\ }\textbf {\bibinfo {volume} {17}},\ \bibinfo {pages}
  {821--825} (\bibinfo {year} {1976})}\BibitemShut {NoStop}
\bibitem [{\citenamefont {Lindblad}(1976)}]{Lindblad1976}
  \BibitemOpen
  \bibfield  {author} {\bibinfo {author} {\bibfnamefont {G.}~\bibnamefont
  {Lindblad}},\ }\bibfield  {title} {\enquote {\bibinfo {title} {On the
  generators of quantum dynamical semigroups},}\ }\href
  {https://doi.org/10.1007/BF01608499} {\bibfield  {journal} {\bibinfo
  {journal} {Commun. Math. Phys.}\ }\textbf {\bibinfo {volume} {48}},\ \bibinfo
  {pages} {119--130} (\bibinfo {year} {1976})}\BibitemShut {NoStop}
\bibitem [{\citenamefont {Breuer}\ and\ \citenamefont
  {Petruccione}(2002)}]{Breuer2002}
  \BibitemOpen
  \bibfield  {author} {\bibinfo {author} {\bibfnamefont {H.-P.}\ \bibnamefont
  {Breuer}}\ and\ \bibinfo {author} {\bibfnamefont {F.}~\bibnamefont
  {Petruccione}},\ }\href
  {https://doi.org/10.1093/acprof:oso/9780199213900.001.0001} {\emph {\bibinfo
  {title} {The Theory of Open Quantum Systems}}}\ (\bibinfo  {publisher}
  {Oxford University Press},\ \bibinfo {address} {Oxford},\ \bibinfo {year}
  {2002})\BibitemShut {NoStop}
\bibitem [{\citenamefont {Bittner}, \citenamefont {Silva-Acuña},\ and\
  \citenamefont {Li}(2026)}]{JCP3}
  \BibitemOpen
  \bibfield  {author} {\bibinfo {author} {\bibfnamefont {E.~R.}\ \bibnamefont
  {Bittner}}, \bibinfo {author} {\bibfnamefont {C.}~\bibnamefont
  {Silva-Acuña}},\ and\ \bibinfo {author} {\bibfnamefont {H.}~\bibnamefont
  {Li}},\ }\href {https://arxiv.org/abs/2606.22530} {\enquote {\bibinfo {title}
  {Liouvillian geometry of multidimensional spectra: Pathway transport and
  observational holonomy in open quantum systems},}\ } (\bibinfo {year}
  {2026}),\ \Eprint {https://arxiv.org/abs/2606.22530} {arXiv:2606.22530
  [quant-ph]} \BibitemShut {NoStop}
\bibitem [{\citenamefont {Avron}\ \emph {et~al.}(2009)\citenamefont {Avron},
  \citenamefont {Fraas}, \citenamefont {Graf},\ and\ \citenamefont
  {Grech}}]{Avron2000}
  \BibitemOpen
  \bibfield  {author} {\bibinfo {author} {\bibfnamefont {J.~E.}\ \bibnamefont
  {Avron}}, \bibinfo {author} {\bibfnamefont {M.}~\bibnamefont {Fraas}},
  \bibinfo {author} {\bibfnamefont {G.~M.}\ \bibnamefont {Graf}},\ and\
  \bibinfo {author} {\bibfnamefont {P.}~\bibnamefont {Grech}},\ }\bibfield
  {title} {\enquote {\bibinfo {title} {Adiabatic response for {Lindblad}
  dynamics},}\ }\href {https://doi.org/10.1007/s00220-008-0673-8} {\bibfield
  {journal} {\bibinfo  {journal} {Commun. Math. Phys.}\ }\textbf {\bibinfo
  {volume} {287}},\ \bibinfo {pages} {651--675} (\bibinfo {year}
  {2009})}\BibitemShut {NoStop}
\bibitem [{\citenamefont {Griewank}\ and\ \citenamefont
  {Walther}(2008)}]{GriewankWalther2008}
  \BibitemOpen
  \bibfield  {author} {\bibinfo {author} {\bibfnamefont {A.}~\bibnamefont
  {Griewank}}\ and\ \bibinfo {author} {\bibfnamefont {A.}~\bibnamefont
  {Walther}},\ }\href@noop {} {\emph {\bibinfo {title} {Evaluating Derivatives:
  Principles and Techniques of Algorithmic Differentiation}}},\ \bibinfo
  {edition} {2nd}\ ed.\ (\bibinfo  {publisher} {SIAM},\ \bibinfo {year}
  {2008})\BibitemShut {NoStop}
\bibitem [{\citenamefont {Baydin}\ \emph {et~al.}(2018)\citenamefont {Baydin},
  \citenamefont {Pearlmutter}, \citenamefont {Radul},\ and\ \citenamefont
  {Siskind}}]{Baydin2018}
  \BibitemOpen
  \bibfield  {author} {\bibinfo {author} {\bibfnamefont {A.~G.}\ \bibnamefont
  {Baydin}}, \bibinfo {author} {\bibfnamefont {B.~A.}\ \bibnamefont
  {Pearlmutter}}, \bibinfo {author} {\bibfnamefont {A.~A.}\ \bibnamefont
  {Radul}},\ and\ \bibinfo {author} {\bibfnamefont {J.~M.}\ \bibnamefont
  {Siskind}},\ }\bibfield  {title} {\enquote {\bibinfo {title} {Automatic
  differentiation in machine learning: a survey},}\ }\href@noop {} {\bibfield
  {journal} {\bibinfo  {journal} {J. Mach. Learn. Res.}\ }\textbf {\bibinfo
  {volume} {18}},\ \bibinfo {pages} {1--43} (\bibinfo {year}
  {2018})}\BibitemShut {NoStop}
\bibitem [{\citenamefont {Schoenholz}\ and\ \citenamefont
  {Cubuk}(2020)}]{SchoenholzCubuk2021}
  \BibitemOpen
  \bibfield  {author} {\bibinfo {author} {\bibfnamefont {S.~S.}\ \bibnamefont
  {Schoenholz}}\ and\ \bibinfo {author} {\bibfnamefont {E.~D.}\ \bibnamefont
  {Cubuk}},\ }\bibfield  {title} {\enquote {\bibinfo {title} {{JAX}, {M.D.}: A
  framework for differentiable physics},}\ }in\ \href@noop {} {\emph {\bibinfo
  {booktitle} {Advances in Neural Information Processing Systems}}},\
  Vol.~\bibinfo {volume} {33}\ (\bibinfo {year} {2020})\ pp.\ \bibinfo {pages}
  {11428--11441},\ \bibinfo {note} {journal version: J. Stat. Mech.
  \textbf{2021}, 124016. DOI: 10.1088/1742-5468/ac3ae9},\ \Eprint
  {https://arxiv.org/abs/1912.04232} {arXiv:1912.04232} \BibitemShut {NoStop}
\bibitem [{\citenamefont {Craig}, \citenamefont {Ares},\ and\ \citenamefont
  {Gauger}(2024)}]{Craig2024DiffLindblad}
  \BibitemOpen
  \bibfield  {author} {\bibinfo {author} {\bibfnamefont {D.~L.}\ \bibnamefont
  {Craig}}, \bibinfo {author} {\bibfnamefont {N.}~\bibnamefont {Ares}},\ and\
  \bibinfo {author} {\bibfnamefont {E.~M.}\ \bibnamefont {Gauger}},\ }\bibfield
   {title} {\enquote {\bibinfo {title} {Differentiable master equation solver
  for quantum device characterization},}\ }\href
  {https://doi.org/10.1103/PhysRevResearch.6.043175} {\bibfield  {journal}
  {\bibinfo  {journal} {Phys. Rev. Res.}\ }\textbf {\bibinfo {volume} {6}},\
  \bibinfo {pages} {043175} (\bibinfo {year} {2024})}\BibitemShut {NoStop}
\bibitem [{\citenamefont {Bradbury}\ \emph {et~al.}(2018)\citenamefont
  {Bradbury}, \citenamefont {Frostig}, \citenamefont {Hawkins}, \citenamefont
  {Johnson}, \citenamefont {Katariya}, \citenamefont {Leary}, \citenamefont
  {Maclaurin}, \citenamefont {Necula}, \citenamefont {Paszke}, \citenamefont
  {VanderPlas}, \citenamefont {Wanderman-Milne},\ and\ \citenamefont
  {Zhang}}]{jax2018github}
  \BibitemOpen
  \bibfield  {author} {\bibinfo {author} {\bibfnamefont {J.}~\bibnamefont
  {Bradbury}}, \bibinfo {author} {\bibfnamefont {R.}~\bibnamefont {Frostig}},
  \bibinfo {author} {\bibfnamefont {P.}~\bibnamefont {Hawkins}}, \bibinfo
  {author} {\bibfnamefont {M.~J.}\ \bibnamefont {Johnson}}, \bibinfo {author}
  {\bibfnamefont {Y.}~\bibnamefont {Katariya}}, \bibinfo {author}
  {\bibfnamefont {C.}~\bibnamefont {Leary}}, \bibinfo {author} {\bibfnamefont
  {D.}~\bibnamefont {Maclaurin}}, \bibinfo {author} {\bibfnamefont
  {G.}~\bibnamefont {Necula}}, \bibinfo {author} {\bibfnamefont
  {A.}~\bibnamefont {Paszke}}, \bibinfo {author} {\bibfnamefont
  {J.}~\bibnamefont {VanderPlas}}, \bibinfo {author} {\bibfnamefont
  {S.}~\bibnamefont {Wanderman-Milne}},\ and\ \bibinfo {author} {\bibfnamefont
  {Q.}~\bibnamefont {Zhang}},\ }\href {https://github.com/jax-ml/jax} {\enquote
  {\bibinfo {title} {{JAX}: Composable transformations of {Python+NumPy}
  programs},}\ } (\bibinfo {year} {2018})\BibitemShut {NoStop}
\bibitem [{\citenamefont {Nakajima}(1958)}]{Nakajima1958}
  \BibitemOpen
  \bibfield  {author} {\bibinfo {author} {\bibfnamefont {S.}~\bibnamefont
  {Nakajima}},\ }\bibfield  {title} {\enquote {\bibinfo {title} {On quantum
  theory of transport phenomena},}\ }\href {https://doi.org/10.1143/PTP.20.948}
  {\bibfield  {journal} {\bibinfo  {journal} {Prog. Theor. Phys.}\ }\textbf
  {\bibinfo {volume} {20}},\ \bibinfo {pages} {948--959} (\bibinfo {year}
  {1958})}\BibitemShut {NoStop}
\bibitem [{\citenamefont {Zwanzig}(1960)}]{Zwanzig1960}
  \BibitemOpen
  \bibfield  {author} {\bibinfo {author} {\bibfnamefont {R.}~\bibnamefont
  {Zwanzig}},\ }\bibfield  {title} {\enquote {\bibinfo {title} {Ensemble method
  in the theory of irreversibility},}\ }\href
  {https://doi.org/10.1063/1.1731409} {\bibfield  {journal} {\bibinfo
  {journal} {J. Chem. Phys.}\ }\textbf {\bibinfo {volume} {33}},\ \bibinfo
  {pages} {1338--1341} (\bibinfo {year} {1960})}\BibitemShut {NoStop}
\bibitem [{\citenamefont {Tanimura}\ and\ \citenamefont
  {Kubo}(1989)}]{Tanimura1989}
  \BibitemOpen
  \bibfield  {author} {\bibinfo {author} {\bibfnamefont {Y.}~\bibnamefont
  {Tanimura}}\ and\ \bibinfo {author} {\bibfnamefont {R.}~\bibnamefont
  {Kubo}},\ }\bibfield  {title} {\enquote {\bibinfo {title} {Time evolution of
  a quantum system in contact with a nearly {Gaussian--Markoffian} noise
  bath},}\ }\href {https://doi.org/10.1143/JPSJ.58.101} {\bibfield  {journal}
  {\bibinfo  {journal} {J. Phys. Soc. Japan}\ }\textbf {\bibinfo {volume}
  {58}},\ \bibinfo {pages} {101--114} (\bibinfo {year} {1989})}\BibitemShut
  {NoStop}
\bibitem [{\citenamefont {Ishizaki}\ and\ \citenamefont
  {Fleming}(2009)}]{Ishizaki2009}
  \BibitemOpen
  \bibfield  {author} {\bibinfo {author} {\bibfnamefont {A.}~\bibnamefont
  {Ishizaki}}\ and\ \bibinfo {author} {\bibfnamefont {G.~R.}\ \bibnamefont
  {Fleming}},\ }\bibfield  {title} {\enquote {\bibinfo {title} {Unified
  treatment of quantum coherent and incoherent hopping dynamics in electronic
  energy transfer: Reduced hierarchy equation approach},}\ }\href
  {https://doi.org/10.1063/1.3155372} {\bibfield  {journal} {\bibinfo
  {journal} {J. Chem. Phys.}\ }\textbf {\bibinfo {volume} {130}},\ \bibinfo
  {pages} {234111} (\bibinfo {year} {2009})}\BibitemShut {NoStop}
\end{thebibliography}

\end{document}